\documentclass[notitlepage,nofootinbib,
superscriptaddress,showpacs,preprintnumbers,amsmath,amssymb,prd,twocolumn,12 point]{revtex4-2}

\usepackage{graphicx}
\usepackage{dcolumn}
\usepackage{bm}
\usepackage{hyperref}
\usepackage[mathlines]{lineno}
\usepackage{float}
\usepackage[T1]{fontenc}
\usepackage{amsmath}
\usepackage{subfigure}
 \DeclareUnicodeCharacter{202F}{FIX ME!!!!}
   \DeclareUnicodeCharacter{2212}{-}

\begin{document}


\title{The $\mu-\tau$ Counter Reflection Symmetry}

\author{Pralay Chakraborty}%
 \email{pralaychakraborty8@gmail.com}
\affiliation{Gauhati University, India}%

\author{Manash Dey}%
 \email{manashdey1272@gmail.com}
\affiliation{Maibang Degree College, India}%



\date{\today}

\begin{abstract}

A novel symmetry for the neutrino mass matrix is proposed that naturally accommodates an inverted mass hierarchy while offering additional phenomenological advantages. The corresponding texture can be realized within a minimal framework based on $\Delta(27)$ symmetry. 

\end{abstract}

\maketitle


\newcolumntype{P}[1]{>{\centering\arraybackslash}p{#1}}

 Neutrino oscillation experiments have achieved precision measurements of the mixing angles\,($\theta_{12}, \theta_{13}, \theta_{23}$) and mass-squared differences\,($\Delta m_{21}^2$ and $\Delta m_{31}^2$)\,\cite{nufit}. Recently, the JUNO experiment has achieved remarkable precision in the measurement of $\theta_{12}$ and $\Delta m_{21}^2$\,\cite{JUNO:2025gmd}. However, the octant of $\theta_{23}$ and the precise range of the Dirac CP phase $\delta$ remain undetermined. In addition, important quantities such as the two Majorana phases\,($\alpha, \beta$) and the individual neutrino mass eigenvalues\,($m_1, m_2, m_3$) are not accessible through these experiments. In this context, model builders explore different approaches to understand the neutrino mass matrix, as it accommodates all nine physical observables. The predictions for these observables are manifested through correlations among the matrix elements. Constructing a predictive neutrino mass matrix with a minimal number of parameters and deriving it from a robust theoretical framework is undoubtedly a challenging task. Several phenomenological concepts, such as $\mu-\tau$ symmetry\,\cite{Altarelli:2005yx}, texture zeroes\,\cite{Xing:2002ta}, hybrid textures\,\cite{Liu:2013oxa}, and vanishing minors\,\cite{Lashin:2009yd}, have been extensively explored in the literature. However, current experimental data have ruled out exact $\mu-\tau$ symmetry, as it predicts a vanishing $\theta_{13}$\,\cite{nufit}. In this regard, deviations from exact $\mu-\tau$ symmetry in the form of $\mu-\tau$ reflection symmetry\,\cite{Xing:2022uax}, $\mu-\tau$ antisymmetry\,\cite{Xing:2020ijf}, and mixed $\mu-\tau$ symmetry\,\cite{Dey:2022qpu, Chakraborty:2023msb} have been extensively investigated.

This motivation leads us to propose a new symmetry for the neutrino mass matrix, termed as the $\mu-\tau$ \textit{counter reflection} symmetry, characterized by four independent complex parameters as shown in the following.

\begin{equation}
\label{Mnu}
M_\nu = \begin{bmatrix}
a & b & -b^* \\
 b & c & d\\
 -b^* & d & -c^*
\end{bmatrix},
\end{equation}
 
highlighting two unique correlations that have not been explored earlier in the literature. The proposed texture is consistent with the present neutrino oscillation data and remains testable in future experiments. The proposed texture exhibits several interesting features. For instance, the texture naturally excludes the normal hierarchy of neutrino masses while predicting the three neutrino mass eigenvalues. In addition, it constrains $\theta_{23}$ and the three CP phases within sharp bounds. At first glance, the mass matrix shown in Eq.\,(\ref{Mnu}) may appear arbitrary, and therefore it is important to identify its novel theoretical origin. Interestingly, we observe that the proposed texture can be realized exactly within a type-I seesaw mechanism and dimension-6 based framework starting from $\Delta\,(27)$ symmetry. 

\begin{figure*}
  \centering
    \subfigure[]{\includegraphics[width=0.31\textwidth]{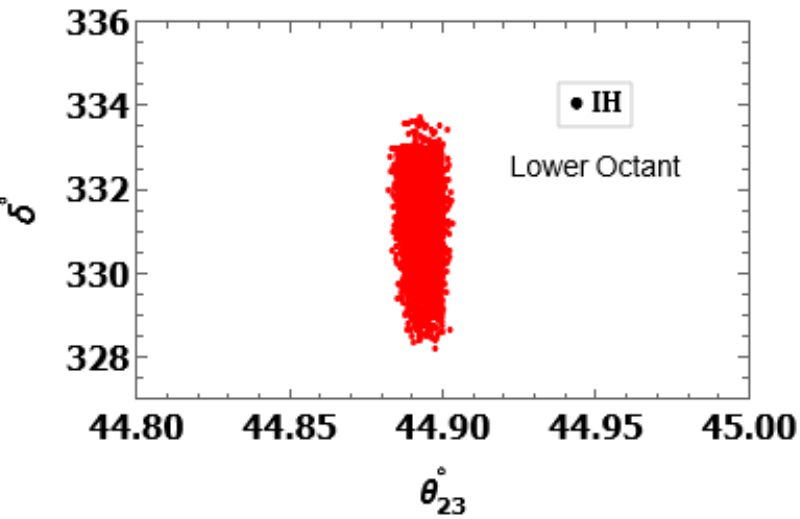}\label{fig:1a}} 
    \subfigure[]{\includegraphics[width=0.3\textwidth]{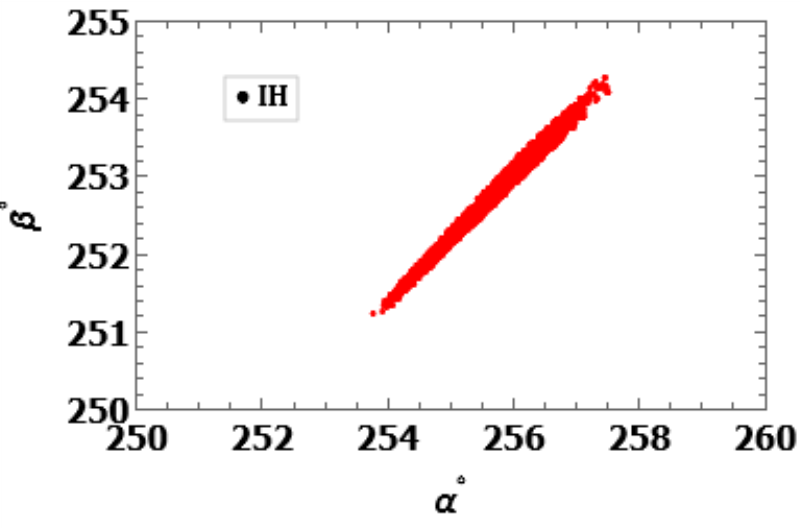}\label{fig:1b}}
    \subfigure[]{\includegraphics[width=0.3\textwidth]{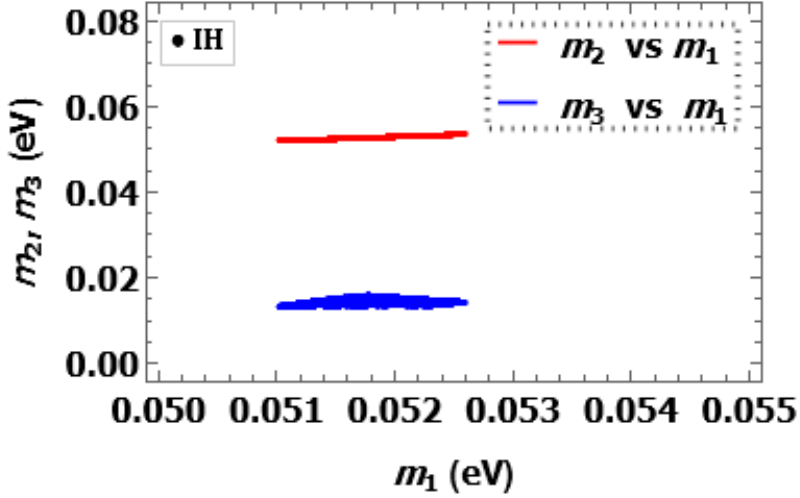}\label{fig:1c}} 
    \subfigure[]{\includegraphics[width=0.31\textwidth]{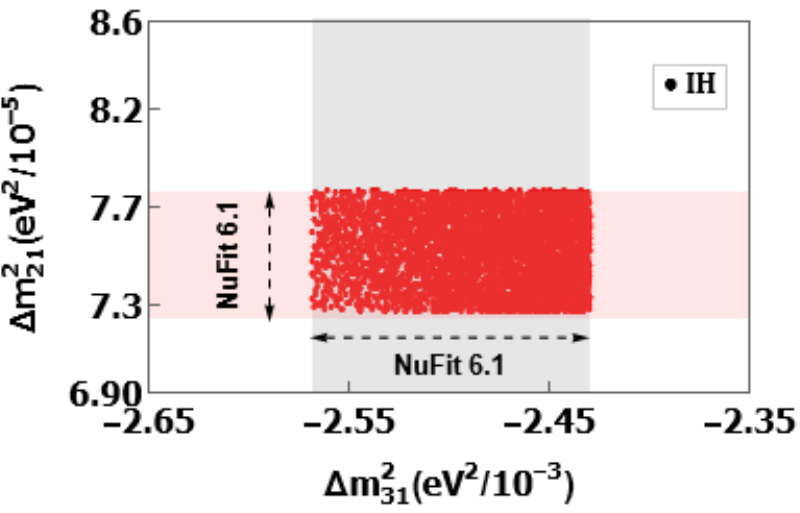}\label{fig:1d}}
    \subfigure[]{\includegraphics[width=0.32\textwidth]{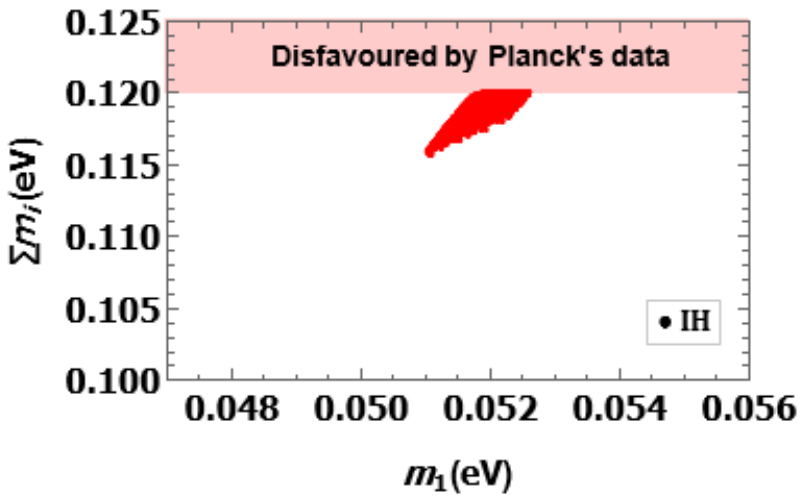}\label{fig:1e}} 
    \subfigure[]{\includegraphics[width=0.3\textwidth]{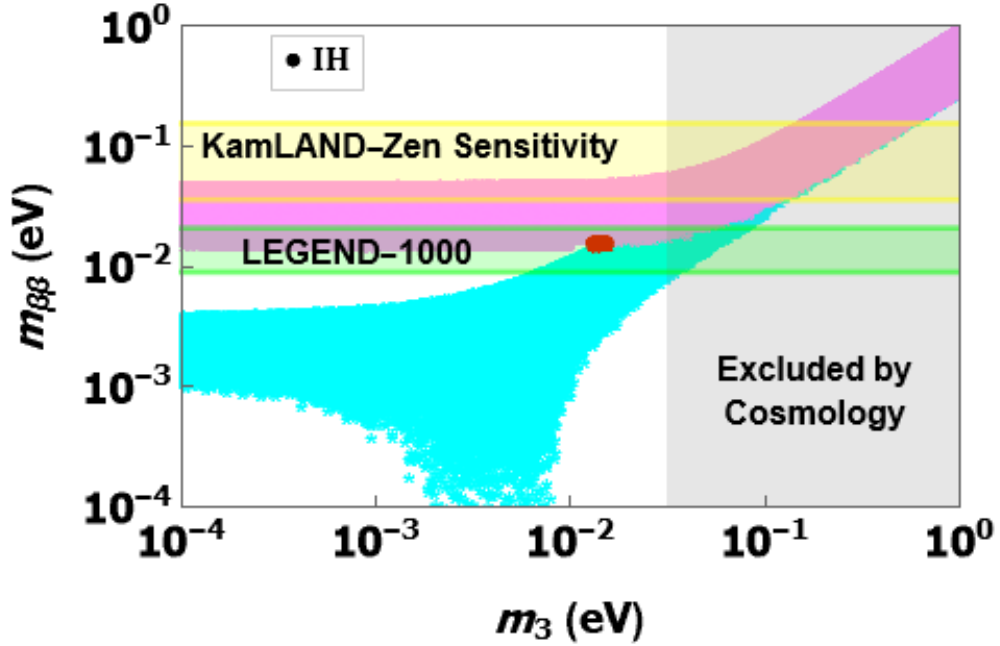}\label{fig:1f}} 
   \caption{ The correlation plots between (a) $\theta_{23}$ vs $\delta$, (b) $\alpha$ vs $\beta$, (c) $m_1$ and $m_2$ vs $m_3$, (d) $\Delta\,m_{21}^2$ vs $\Delta\,m_{31}^2$, (e) $\sum m_i$ vs $m_1$, (f) The effective Majorana neutrino mass $m_{\beta\beta}$ vs $m_1$. }
\label{fig:mass parameters}
\end{figure*}

\begin{figure*}
  \centering
    \subfigure[]{\includegraphics[width=0.32\textwidth]{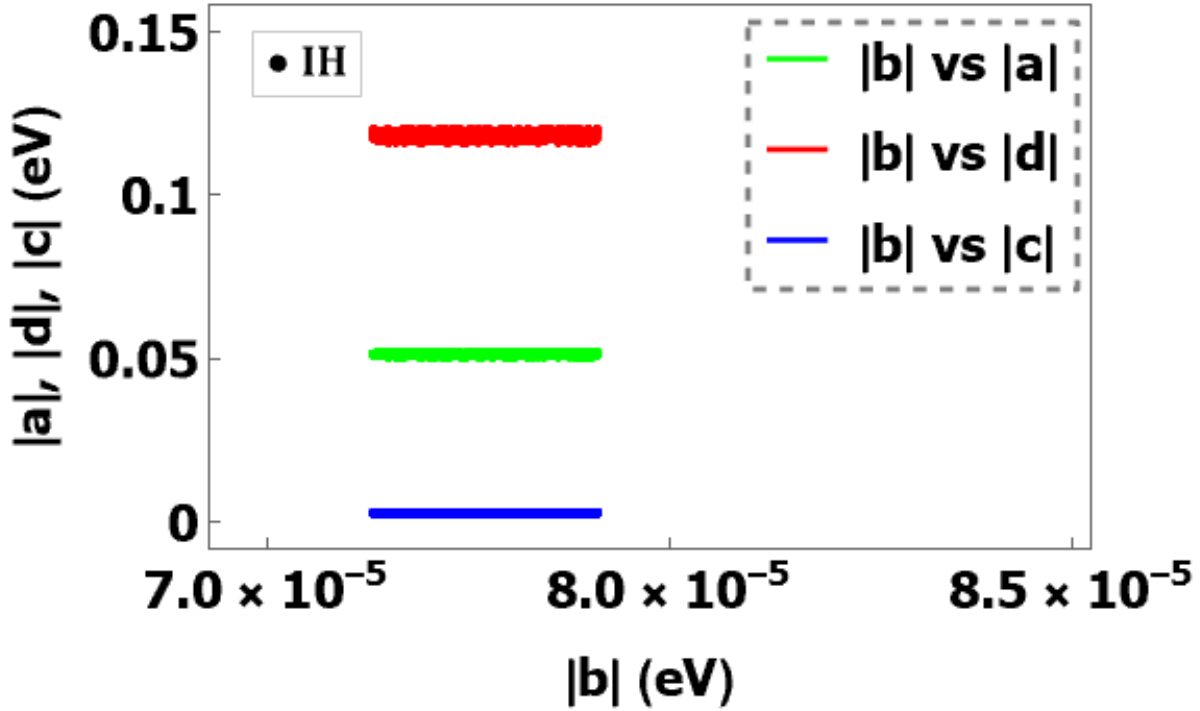}\label{fig:3a}}
    \subfigure[]{\includegraphics[width=0.3\textwidth]{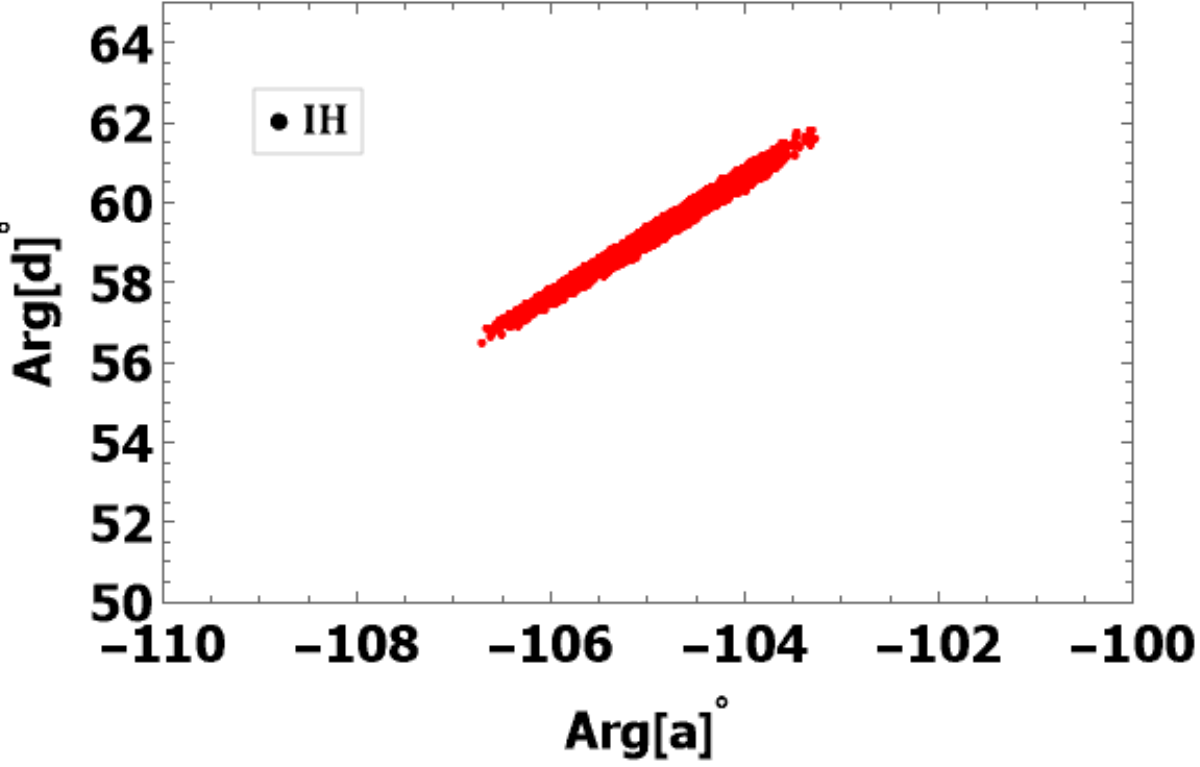}\label{fig:3b}} 
    \subfigure[]{\includegraphics[width=0.32\textwidth]{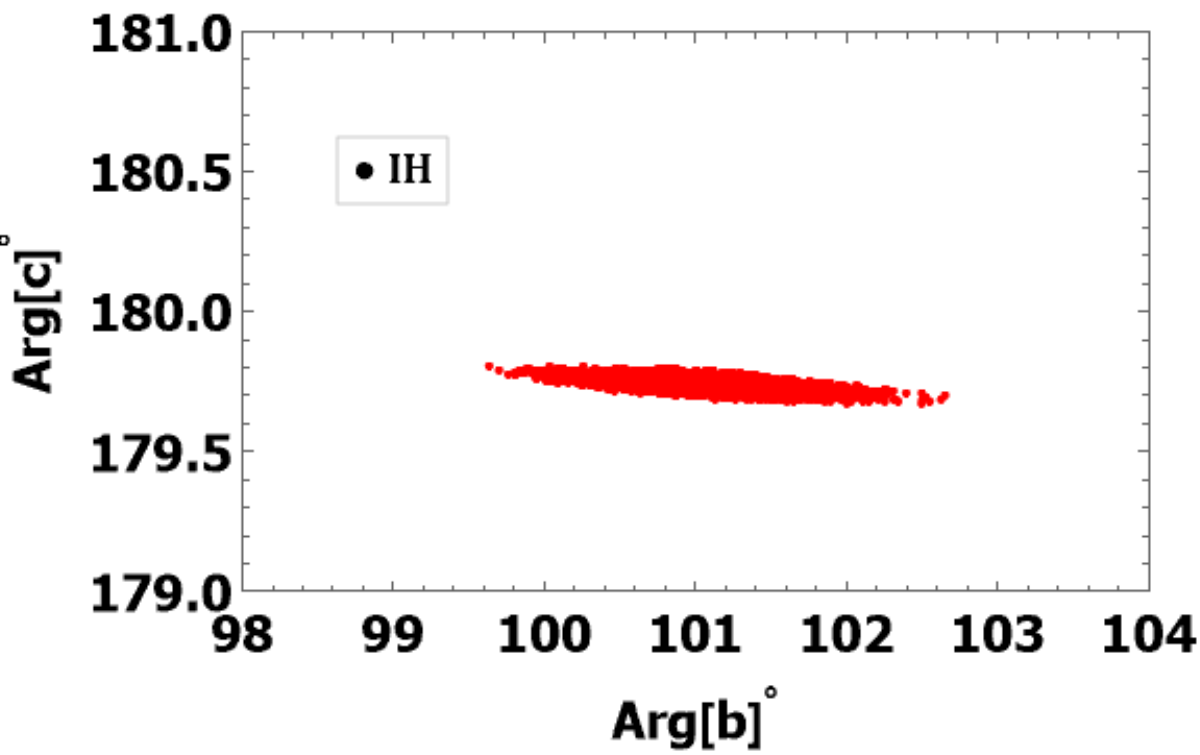}\label{fig:3c}} 
   \caption{ The correlation plots between (a) $|a|$, $|d|$, $|c|$ vs $|b|$, (b) $\text{Arg}[d]$ vs $\text{Arg}[a]$, (c) $\text{Arg}[c]$ vs $\text{Arg}[b]$. }
\label{fig:texture parameters}
\end{figure*}

We extract the information of the observable parameters from $M_\nu$. Interestingly, the texture resolves the octant degeneracy of $\theta_{23}$ by predicting it to lie around $45^{\circ}$ in the lower octant. Similarly, the Dirac CP phase $\delta$ is predicted within the fourth quadrant with considerable precision. On the other hand, the Majorana phases $\alpha$ and $\beta$ are constrained within the third quadrant. The proposed mass matrix excludes the possibility of a normal hierarchy of neutrino masses and further disfavours the extreme scenario of vanishing $m_3$. We emphasize that the predicted values of $\Delta,m_{21}^2$ and $\Delta,m_{31}^2$ remain consistent with the experimental $3\sigma$ bounds\,\cite{nufit}, while $\sum m_i$ lies below the cosmological upper bound of $0.12$ eV\,\cite{Planck:2018vyg}. The corresponding graphical analysis is presented in Fig\,(\ref{fig:1a})-(\ref{fig:1e}) and the minimum and maximum values of the studied parameters are listed in Table\,(\ref{physical parameters}). As mentioned earlier, $\alpha$ and $\beta$ are not directly observable quantities. However, their presence significantly influences the effective Majorana neutrino mass, $m_{\beta\beta}$\,\cite{Schechter:1981bd}. Consequently, as a potential implication of Eq.\,(\ref{Mnu}), we obtain $m_{\beta\beta}$ in the range $(25.7-28.97)$ meV (see Fig.\,(\ref{fig:1f})), which lies within the sensitivity reach of LEGEND-1000\,\cite{LEGEND:2021bnm}. A graphical visualization of the texture parameters is presented in Fig.\,(\ref{fig:texture parameters}).

\begin{table}[b]
\centering
\begin{tabular*}{0.47\textwidth}{@{\extracolsep{\fill}} ccc}
\hline
\hline
Parameter & Minimum Value &  Maximum Value \\
\hline
\hline
$\theta_{23} /^\circ$ & 44.88 & 44.90 \\
\hline
$\delta /^\circ$ & 328.19 & 333.69 \\
\hline
$\alpha/^\circ$ & 253.76  & 257.51  \\
\hline
$\beta/^\circ$ & 251.24 & 254.27  \\
\hline
$m_1$/eV & 0.051 & 0.052 \\
\hline
$m_2$/eV & 0.052 & 0.053  \\
\hline
$m_3$/eV & 0.013 & 0.015  \\
\hline
\end{tabular*}
\caption{Shows the maximum and minimum values of  $\theta_{23}$, $\delta$, $\alpha$, $\beta$, $m_1$, $m_2$ and $m_3$ predicted from the proposed texture.}
\label{physical parameters}
\end{table}
 
\begin{figure*}
  \centering
    \subfigure[]{\includegraphics[width=0.32\textwidth]{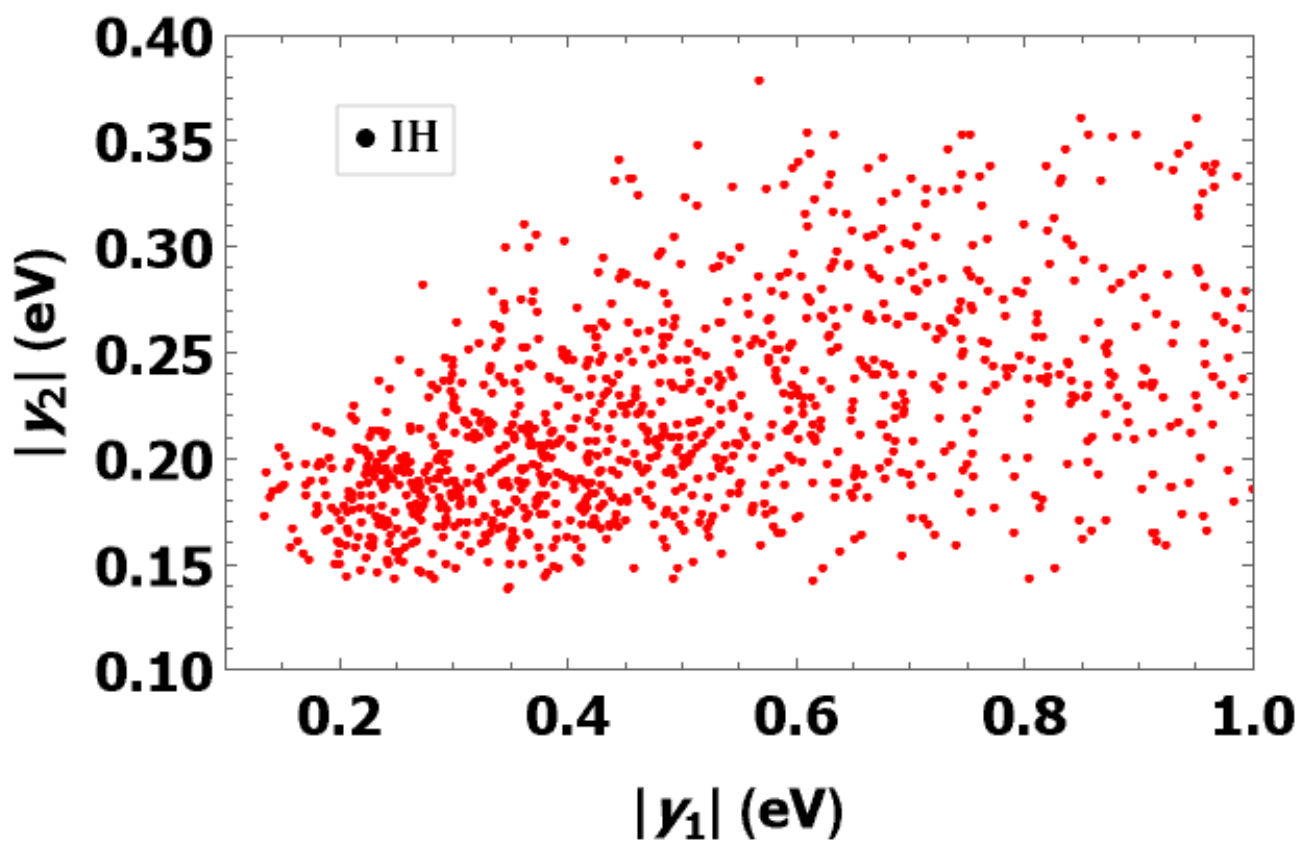}\label{fig:3a}}
    \subfigure[]{\includegraphics[width=0.31\textwidth]{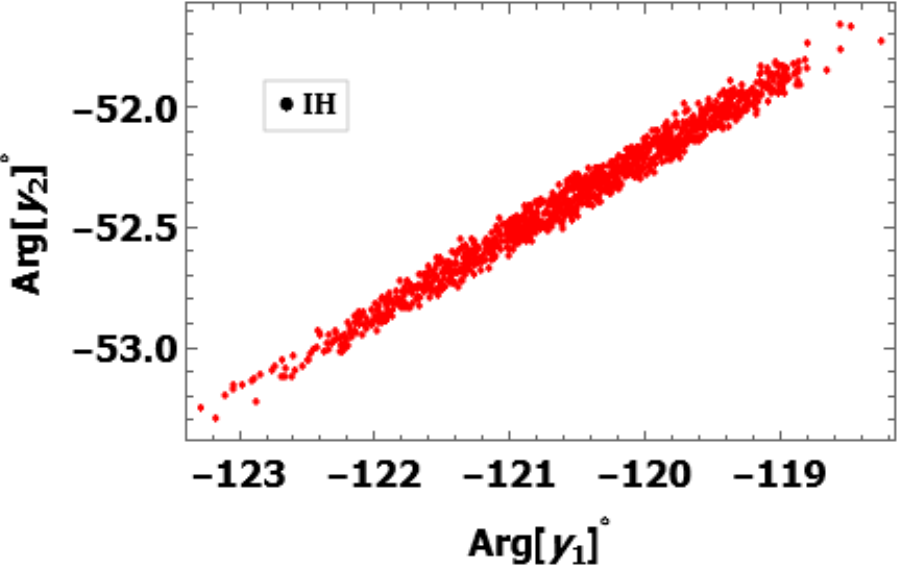}\label{fig:3b}} 
    \subfigure[]{\includegraphics[width=0.3\textwidth]{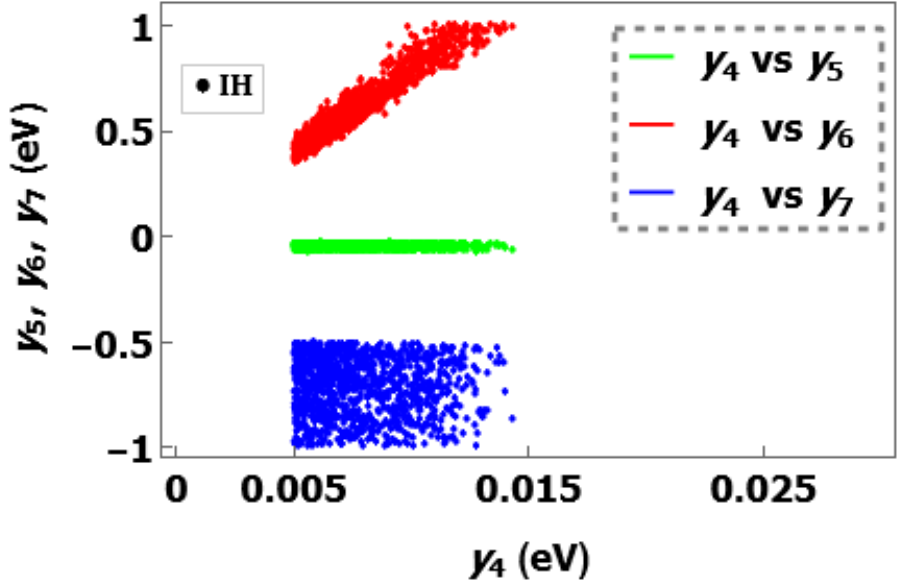}\label{fig:3c}} 
   \caption{ The parameter space of the Yukawa couplings. }
\label{fig:model parameters}
\end{figure*}

In order to understand the possible origin of the posited texture, we consider a framework equipped with type-I seesaw and two dimension-6 operators. The SM group is extended by $SU(2)_L \times U(1)_Y \times \Delta(27) \times Z_{7}$ along with newly added scalar fields and right-handed neutrinos (see Table\,(\ref{Field Content of M})). The effective Lagrangian is shown below,   

\begin{align}
- \mathcal{L}_Y &= \frac{y_{e}}{\Lambda} (\bar{D}_{l_{L}}\phi)H e_{R} + \frac{ y_{\mu}}{\Lambda}(\bar{D}_{l_{L}}\phi)H\mu_{R} + \frac{ y_{\tau}}{\Lambda}(\bar{D}_{l_{L}}\phi)H \tau_{R}\nonumber\\&+ y_{1}(\bar{D}_{l_{L}}\phi )\tilde{H}\nu_{eR}+\frac{y_2}{\Lambda}(\bar{D}_{l_{L}}\phi)\tilde{H} \nu_{\mu_R} + \frac{y_3}{\Lambda} (\bar{D}_{l_{L}}\phi)\tilde{H}\nonumber\\&\nu_{\tau_R}+ M_1 (\bar{\nu}^c_{\mu_{R}}\nu_{\mu_{R}}) +M_2(\bar{\nu}^c_{e_{R}}\nu_{\tau_{R}} + \bar{\nu}^c_{\tau_{R}}\nu_{e_{R}})+ \frac{y_4'}{\Lambda^2} \nonumber\\&(\bar{D}_{l_{L}} D_{l_{L}}^{c})_{3^*_{s1}} \chi H^2 +\frac{y_{5}}{\Lambda^2}(\bar{D}_{l_{L}} D_{l_{L}}^{c})_{3^*_{s2}} \chi H^2+\frac{y_{6}}{\Lambda^2}(\bar{D}_{l_{L}}\nonumber\\& D_{l_{L}}^{c})_{3^*_{s1}} \psi^* H^2 +\frac{y_7'}{\Lambda^2}(\bar{D}_{l_{L}} D_{l_{L}}^{c})_{3^*_{s2}} \psi^* H^2+ h.c.,
\label{Yukawa Lagrangian M1}
\end{align}

where, $\Lambda$ defines the cut-off scale of the theory. The presence of the auxiliary group $Z_7$ in the model restricts the undesirable terms that are allowed by $\Delta(27)$. Achieving the desired texture depends on the proper vacuum expectation values\,(vev) alignment of the scalar fields. We consider: $\langle H \rangle_{0}=v_{H}$, $\langle \phi \rangle_{0}=v_{\phi}(1,0,0)^{T}$, $\langle \chi \rangle_{0}=v_{\chi}(0,1,1)^{T}$ and $\langle \psi \rangle_{0}=v_{\psi}(0,1,-1)^{T}$ respectively.  Achieving the desired texture depends on additional constraints. For the present model, $y_{1,2,3}$ are complex and $y_{5,6}$ are real. On the other hand, $y_4'=i y_4$ and $y_7'=i y_7$ are purely imaginary. The framework brings forth a diagonal charged lepton mass matrix. However, to achieve the targeted neutrino mass matrix, we include a phase in the left and right handed diagonalizing matrices as shown below,
\begin{equation}
		U_{l_{L}}= \begin{bmatrix}
			e^{i\zeta} &  0 & 0\\
			0 & 1 & 0\\
			0 & 0 & 1\\
		\end{bmatrix},\quad
		U_{l_{R}}=\text{diag}\begin{bmatrix}
			e^{i\zeta}, & 1, & 1
		\end{bmatrix}.
	\end{equation}
	
	\begin{table}
\centering
\begin{tabular*}{0.48\textwidth}{@{\extracolsep{\fill}} ccccc}
\hline
Field & $SU(2)_L$ & $U(1)_Y$ & $\Delta(27)$ & $Z_7$ \\
\hline\hline
$\overline{D}_{l_L}$ 
& 2 & 1 & 3 & 2 \\
\hline
$D^C_{l_L}$ 
& 2 & 1 & 3 & 2 \\
\hline
$l_R$ 
& 1 & $(-2,-2,-2)$ & $(1_{00},1_{01},1_{02})$ & $(0,0,0)$ \\
\hline
$\nu_{l_R}$ 
& 1 & $(0,0,0)$ & $(1_{01},1_{00},1_{02})$ & $(0,0,0)$ \\
\hline
$H$ 
& 2 & 1 & $1_{00}$ & 0 \\
\hline
$\phi$ 
& 1 & 0 & $3^*$ & 5 \\
\hline
$\chi$ 
& 1 & 0 & 3 & 3 \\
\hline
$\psi$ 
& 1 & 0 & $3^*$ & -3 \\
\hline
\end{tabular*}
\caption{Transformation properties of the various fields under the symmetry group $SU(2)_L \times U(1)_Y \times \Delta(27) \times Z_7$.}
\label{Field Content of M}
\end{table}

\begin{figure}[h]
\centering
\includegraphics[scale=0.3]{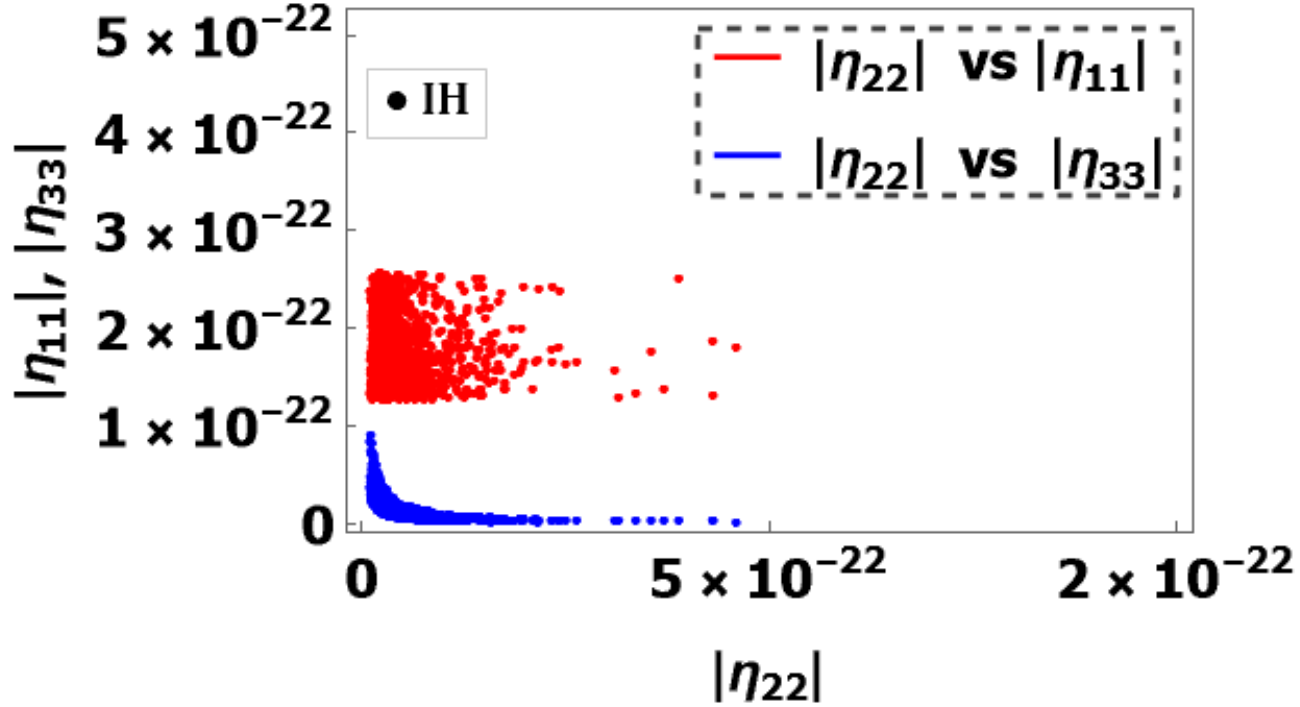}
 \caption{ The prediction of $|\eta_{\alpha\beta}|$ from the present model. It is seen that the nonunitarity effects are highly suppressed.}
\label{fig:nonuni}
\end{figure}

Here, we insist that the phase $\zeta$ appearing in $U_{l_{L}}$ and $U_{l_{R}}$ does not appear directly from the framework itself. We take the liberty to choose the eigenvectors in different possible ways. Needless to mention that the presence $\zeta$ does not alter the texture of $M_l$. However, in the light of charged lepton correction, the $U_{l_L}$ transforms $M_\nu$ in flavour basis and thus $\zeta$ does contribute to the latter. For the present analysis, we fix $\zeta$ at $\pi/2$ and obtain the proposed texture in Eq.\,(\ref{Mnu}) with,

\begin{align}
M_\nu=\begin{bmatrix}
 \frac{v^2 y_2^2 v_{\phi }^2}{\Lambda ^2 M_1} & \frac{y_7 v_{\psi }}{2}+\frac{i v^2 y_5 v_{\chi }}{2 \Lambda ^2} & -\frac{y_7 v_{\psi }}{2}+\frac{i v^2 y_5 v_{\chi }}{2 \Lambda ^2} \\
 \frac{y_7 v_{\psi }}{2}+\frac{i v^2 y_5 v_{\chi }}{2 \Lambda ^2} & y_6 v_{\psi }+\frac{i v^2 y_4 v_{\chi }}{\Lambda ^2} & -\frac{v^2 y_1 y_3 v_{\phi }^2}{\Lambda ^2 M_2} \\
 -\frac{y_7 v_{\psi }}{2}+\frac{i v^2 y_5 v_{\chi }}{2 \Lambda ^2} & -\frac{v^2 y_1 y_3 v_{\phi }^2}{\Lambda ^2 M_2} & -y_6 v_{\psi }+\frac{i v^2 y_4 v_{\chi }}{\Lambda ^2} \\
\end{bmatrix}
\end{align}

 It is important to have an estimate of the possible domain of the model parameters. The parameter space of the Yukawa couplings is shown in Fig.\,(\ref{fig:model parameters}). To obtain these viable conditions, we stick to the following set of parameter values: $\Lambda \sim 10^{14}$\,GeV, $v_\phi \sim 10^{13}$\,GeV, $v_\psi \sim 0.05$\,eV, and $v_\chi \sim 10^{13}$\,GeV respectively. Needless to mention, $v_H$ is fixed at 246\,GeV.

In our model, the charged Higgs scalar field, $H^+$, is a would-be Goldstone boson that is ultimately eaten by  $W^+$, via the Higgs mechanism. The $W^+$ boson couples a left-handed (LH) charged lepton to a left-handed (LH) neutrino through the interaction Lagrangian. The branching ratio of a charged lepton flavour violation\,(CLFV) process with a light neutrino as a mediator is very low ($BR(\mu \to e \gamma) \approx 10^{-54}$)\,\cite{ParticleDataGroup:2024cfk}. However, in the Type-I seesaw framework, the left-handed neutrinos, $\nu_L$, can mix with the heavy right-handed Majorana neutrinos through the mixing matrix $\Theta \approx M_D M_R^{-1}$, where $M_D$ and $M_R$ denote the Dirac-type and heavy right-handed neutrino mass matrices, respectively. Neglecting the highly suppressed contribution from the light neutrinos and retaining only the heavy-neutrino contribution, the approximate expression can be written as follows,
\begin{align}
\label{eqlalbg}
\mathrm{BR}(\ell_\alpha \to \ell_\beta \gamma)
\approx
\frac{3\alpha}{32\pi}
\left|
\sum_i
\Theta_{\alpha i}\,
\Theta^*_{\beta i}\,
F\!\left(\frac{M_i^2}{M_W^2}\right)
\right|^2,
\end{align}

where $M_i$ is the mass of $i$-th heavy neutrino and the function, $F(x) \approx4/3$ for $M_i >> M_W$.  With $M_D,\,M_R$ and $U_L$ in our model, we obtain $\Theta \Theta^\dagger$ in flavour basis as shown below,

\begin{align}
\Theta \Theta^\dagger \approx 
\begin{bmatrix}
 \frac{v^2 v_\phi^2 y_2^2}{\Lambda^2 M_1^2} & 0 & 0 \\
 0 & \frac{v^2 v_\phi^2 y_1^2}{\Lambda^2 M_2^2} & 0 \\
 0 & 0 & \frac{v^2 v_\phi^2 y_3^2}{\Lambda^2 M_2^2}
\end{bmatrix}
\end{align}

Since gauge boson mediated CLFV amplitudes are proportional to the off-diagonal entries of ($\Theta\Theta^\dagger$), all such contributions vanish identically in the present framework. Therefore, the branching ratios for processes such as ($\mu\to e\gamma$), ($\mu\to 3e$), ($\tau\to e\gamma$), ($\tau\to\mu\gamma$), and related gauge-mediated CLFV channels are highly suppressed. Thus, the heavy-light neutrino mixing does not induce flavour transitions among charged leptons.

In general, additional neutral fermionic states in seesaw models can lead to deviations from the exact unitarity of the leptonic mixing matrix. The nonunitarity effects originate from the mixing between the active neutrinos and the heavy singlet fermions. As a consequence of this mixing, the effective leptonic mixing matrix associated with the light neutrinos deviates from exact unitarity and can be written as,

\begin{equation}
N \simeq \left(I-\frac{1}{2}\Theta\Theta^\dagger\right)U_{\rm PMNS},
\end{equation}
where, $U_{\rm PMNS}$ denotes the unitary PMNS matrix in the absence of heavy-light mixing. The corresponding nonunitarity parameter is therefore defined as
\begin{equation}
\eta = \frac{1}{2}\Theta\Theta^\dagger.
\end{equation}

The deviations are tightly constrained by electroweak precision data, neutrino oscillation experiments and weak decay observables: $|\eta_{11}| < 5.5 \times 10^{-3}$, $|\eta_{22}| < 5.0 \times 10^{-3}$ and $|\eta_{33}| < 5.0 \times 10^{-3}$ respectively\,\cite{Fernandez-Martinez:2007iaa}. From our model, we investigate the variation of the parameter $\eta_{\alpha \beta}$ and it is found that the nonunitarity effects are highly suppressed\,(see Fig.\,(\ref{fig:nonuni})).

In summary, we have proposed a new symmetry for the neutrino mass matrix, termed as the $\mu-\tau$ counter reflection symmetry, with four complex parameters. The texture has accounted for all observed neutrino parameters, has ruled out the normal mass hierarchy by predicting the individual neutrino mass eigenvalues, and has precisely predicted $\theta_{23}$ and $\delta$, placing them in the lower octant and fourth quadrant, respectively. In addition, the texture has estimated the Majorana phases to lie in the third quadrant. This has been particularly relevant in light of the sensitivity of ongoing and future experiments. For $\theta_{12}$ and $\theta_{13}$, the model has made no specific predictions. However, it has remained consistent with the corresponding experimental $3\sigma$ bounds. Furthermore, we have derived the neutrino mass matrix in its exact form, maintaining the independence of the parameters, within a framework encompassing the type-I seesaw mechanism and two dimension-6 operators under the $SU(2)_L \times U(1)_Y \times \Delta(27) \times Z_{7}$ symmetry. To explore the phenomenological implications of the model, we have further extended our discussion to $m_{\beta\beta}$, CLFV, and the nonunitarity of the lepton mixing matrix.

The CLFV processes originating from the charged lepton sector can be mediated by physical CP even and CP odd scalars and may lead to sizeable contributions. In addition, it would be worthwhile to investigate the stability of the proposed texture under renormalization group evolution. Both of these aspects, however, require a comprehensive analysis of the scalar sector. Such an investigation lies beyond the scope of the present work and is therefore left for future study.

\section{Acknowledgement}

The authors thank S. Roy and S.T. Goswami, Gauhati University for fruitful discussions.

\bibliography{ref.bib}

\end{document}